# Approximating Accelerator Impedances with Resonator Networks

Brian J. Vaughn*

*Fermi National Accelerator Laboratory, Batavia, Illinois 60510*

It is common in the accelerator community to use the impedance of accelerator components to describe wake interactions in the frequency domain. However, it is often desirable to understand such wake interactions in the time domain in a general manner for excitations that are not necessarily Gaussian in nature. The conventional method for doing this involves taking the inverse Fourier Transform of the component impedance, obtaining the Green's Function, and then convolving it with the desired excitation distribution. This method can prove numerically cumbersome, for a convolution integral must be evaluated for each individual point in time when the wake function is desired. An alternative to this method would be to compute the wake function analytically, which would sidestep the need for repetitive integration. Only a handful of cases, however, are simple enough for this method to be tenable. One of these cases is the case where the component in question is an RLC resonator, which has a closed-form analytical wake function solution. This means that a component which can be represented in terms of resonators can leverage this solution. As it happens, common network synthesis techniques may be used to map arbitrary impedance profiles to RLC resonator networks in a manner the accelerator community has yet to take advantage of. In this work, we will use Foster Canonical Resonator Networks and partial derivative descent optimization to develop a technique for synthesizing resonator networks that well approximate the impedances of real-world accelerator components. We will link this synthesis to the closed-form resonator wake function solution, giving rise to a powerful workflow that may be used to streamline beam dynamics simulations.

## 1. Introduction

Wake field and impedance modeling of accelerator components is a critical aspect of the accelerator design process given the strong electromagnetic interactions between the particle beam and its surroundings. Because of the implications such interactions have on the beam's stability and dynamic evolution over time, much effort has been devoted to the electromagnetic simulation of accelerator components, resulting in the development and use of numerical codes such as CST Particle Studio [1], GdfidL, [2], Superfish [3], etc. While these codes are powerful in their ability to model E/M fields, they do not typically output the Green's Function of an accelerator component. That is, they do not give as an output the response of the component to a single moving point charge (impulse current), which would allow for the response of an arbitrary charge distribution to be found via convolution of the impulse current response with the greater distribution. Instead, the primary output is typically the wake function of a specified Gaussian excitation bunch distribution or the frequency-domain impedance of the component, which may be moved to the time-domain via an inverse Fourier transform and convolved with an arbitrary excitation bunch current to obtain a wake function [4]. Restricting ourselves to the longitudinal regime, we may represent this formalism in terms of a longitudinal wake voltage $V(t)$, a bunch excitation current $I(t)$, and an accelerator component impulse response (Green's Function) $z(t)$, all with respective Fourier Transforms $\tilde{V}(\omega), \tilde{I}(\omega)$, and $\tilde{Z}(\omega)$. To simulate time-dependent beam dynamics, we are often interested in $V(t)$, which is the potential (work done per unit charge) experienced by a particle that enters the wake field of an accelerator component at time $t$ excited by a bunch represented by $I(t)$. The relationships between these quantities are as follows:

$$V(t) = I(t) * z(t) = \int_{-\infty}^{\infty} I(t-\tau)z(\tau)d\tau, \tag{1a}$$

$$\tilde{V}(\omega) = \tilde{I}(\omega)\tilde{Z}(\omega). \tag{1b}$$

Again, if $\tilde{Z}(\omega)$ is the output of the simulation, the only way to obtain $z(t)$ is to conduct an inverse Fourier transform. Furthermore, even if $z(t)$ is on hand, the numerical computation of $V(t)$ requires the evaluation of a convolution integral (which must be truncated in the numerical context) for every time point of interest, which can be quite computationally expensive. These factors present a significant inconvenience for many accelerator modeling problems. One way around this obstacle would be to obtain an analytical result for $V(t)$, which would allow for $V(t)$ to be known at every time $t$ without the need to conduct the convolution integral over and over, though an analytical result is difficult to achieve in general. There is, however, a case quite common in the accelerator space where an analytical result is available, namely, the case where the accelerator component in question is a parallel RLC resonator. Recently, in fact, a closed-form expression was derived for the wake potential $V(t)$ over a parallel RLC resonator excited by a bunch current $I(t)$ with Fourier Transform

$\tilde{I}(\omega)$. This expression is quoted below [5]:

$$V(t) = \frac{\omega_r R}{Q}\left(\frac{1}{r^+ - r^-}\right)\left(r^+\tilde{I}(r^+)e^{jr^+t} - r^-\tilde{I}(r^-)e^{jr^-t}\right), \tag{2}$$

$$r^\pm = \frac{\omega_r}{2}\left(j/Q \pm \sqrt{4 - \frac{1}{Q^2}}\right). \tag{3}$$

Where $R$ is the resonator shunt resistance, $\omega_r$ is the resonant frequency, and $Q$ is the quality factor. Note that this expression is truly closed-form, whereas, to the author's knowledge, expressions historically used to compute wake functions of parallel RLC resonators have involved the Error Function, which cannot be expressed with elementary functions [6]. Note that eqn. (2) is valid only if $\tilde{I}(\omega)$ has no poles (i.e., is holomorphic). If this is not the case, then eqn. (17a) of [5] is to be used instead, though that expression will not be quoted here due to its unnecessary complexity regarding the discussion herein. It should also be stated that eqns. (2) and (3) here are slightly different than the equations in [5] (eqns. (5b) and (13a) specifically). This is because the development in [5] invoked the change of variables $u \equiv \frac{\omega}{\omega_r}$ and then made evaluations in the $u$-domain. Here, the Fourier Transform evaluations are made in the $\omega$-domain and the $r^\pm$ definition includes $\omega_r$ as a result.

Due to the simplicity of the expression above, a reasonable strategy for wake potential modeling would be to approximate the impedance of the component in question using RLC resonators. This technique is not new and is a fairly common practice in the accelerator community [4], [7-9]. However, methods for executing this approximation appear to be primarily empirical in nature, and thus limited in the accuracy and sophistication of the resulting resonator networks. In other fields, however, optimization techniques along with utilization of a Foster Canonical Network (FCN) (more on this below) have been leveraged to represent generalized impedance responses in terms of RLC resonators [10]-[15]. Here, we will do the same, developing a partial-derivative-based procedural search algorithm that may be used to map broadband accelerator component data to quasi-equivalent resonator networks. The results of this effort, then, will provide a robust technique for mapping complex impedance responses to the time domain while bypassing the need for repetitive computations at different time points. The outline of the following sections is as follows: first, the FCN will be discussed in more detail. Next, the optimization algorithm used for this study will be described. Finally, examples will be given for the networks obtained when the optimization scheme is applied to the impedance data of actual accelerator components used in real machines today.

## 2. Foster Canonical Networks

In 1924, Ronald Foster proved an impactful network analysis theorem stating that general lossless impedance profiles or arbitrary networks may be described in terms of odd functions of frequency composed of alternating resonances and anti-resonances [10]. From there, it follows that one can synthesize non-unique approximate networks for arbitrary impedances that take one of the two forms shown below [10], [16]:

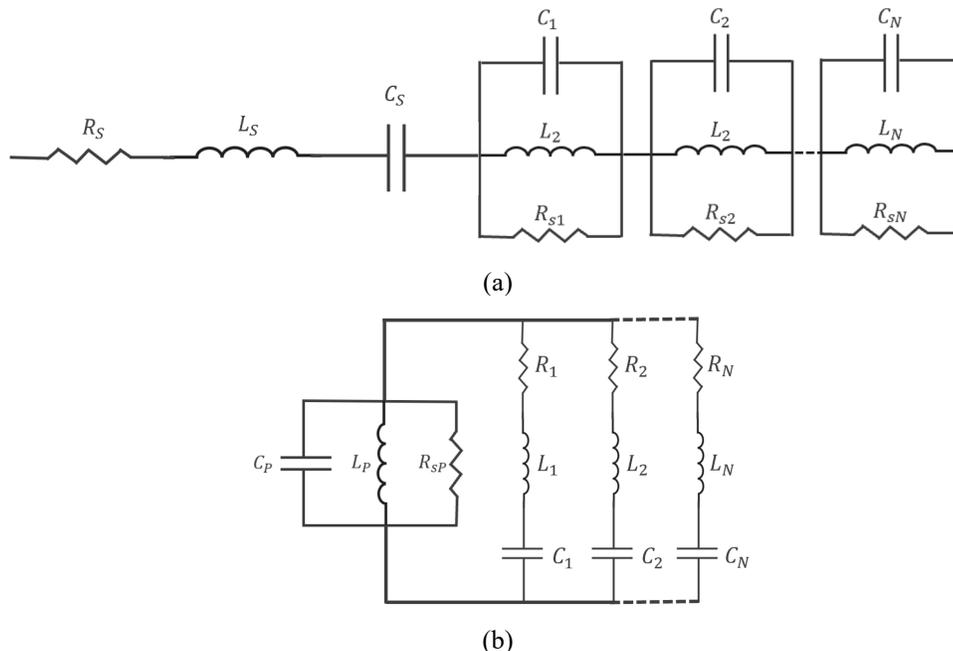

**FIG. 1.** Foster Canonical resonator networks used to synthesize impedance responses. The Form I variant (a) is made primarily of parallel RLC resonators in series with each other as well as one series RLC, whereas Form II (b) consists of series RLC's in parallel with each other as well as one parallel RLC.

As the figure above shows, the canonical forms either consist of series-connected parallel resonators with one series resonator or parallel-connected series resonators with one parallel resonator. Note that the circuits in the figure above contain resistive elements, which were not included in Foster's original formulation. Real passive networks, however, contain loss in general, so the canonical forms above were extended in an intuitive manner by adding resistors to the resonators, thus making their Q's finite without violating Foster's Reactance Theorem. Both forms may be used for network synthesis, but Form I above is far more applicable to accelerator fields. This is because of the interaction between the impedances and the bunch excitation current $I(t)$. If the resonators are in series, then they each experience the same $I(t)$, which means that the total wake potential is merely the summation of the potentials developed over each individual resonator. For that reason, we will only consider Form I going forward.

One aspect of the above Form I circuit that must be developed further is the presence of the series RLC resonator. Since the development in [5] was geared toward RF cavities, only parallel RLC resonators were considered. The applicable theoretical treatment for the case of the series RLC resonator is, however, straightforward. This treatment is detailed thusly. Consider a series RLC resonator with an impedance described by the expression below:

$$Z_s(\omega) = R_s \left[1 - jQ_s \left(\frac{\omega_{rs}}{\omega} - \frac{\omega}{\omega_{rs}}\right)\right], \tag{4}$$

where $R_s$ is the resonator series resistance, $Q_s$ is the resonator quality factor, and $\omega_{rs}$ is the resonant frequency. Since we are interested in the convolution of the above expression's inverse Fourier Transform with the bunch current $\tilde{I}(\omega)$, we are equivalently interested in the inverse Fourier Transform of $\tilde{I}(\omega)Z_s(\omega)$. We may then represent the wake potential developed over the series RLC resonator as

$$V_s(t) = \breve{F}^{-1}\left(R_s\tilde{I}(\omega) + \frac{R_s\omega_{rs}Q_s}{j\omega}\tilde{I}(\omega) + j\omega\frac{R_sQ_s}{\omega_{rs}}\tilde{I}(\omega)\right). \tag{5}$$

From well-documented Fourier Transform rules, we may evaluate this expression by inspection as

$$V_s(t) = R_sI(t) + R_s\omega_{rs}Q_s \int_{-\infty}^{t} I(t)dt + \frac{R_sQ_s}{\omega_{rs}}\frac{dI(t)}{dt}. \tag{6}$$

Invoking eqn. (2), we may now express the total wake potential developed over the Form I FCN:

$$V_{tot}(t) = R_sI(t) + R_s\omega_{rs}Q_s \int_{-\infty}^{t} I(t)dt + \frac{R_sQ_s}{\omega_{rs}}\frac{dI(t)}{dt} + \sum_{1}^{N}\frac{\omega_{rn}R_n}{Q_n}\left(\frac{1}{r_n^+ - r_n^-}\right)\left(r_n^+\tilde{I}(r_n^+)e^{j\omega_{rn}r_n^+t} - r_n^-\tilde{I}(r_n^-)e^{j\omega_{rn}r_n^-t}\right). \tag{7}$$

Note that the $n$ subscripts denote that the parameter belongs to the $n^{th}$ resonator in the parallel RLC chain. Note also that the number of parallel resonators $N$ depends on the problem at hand. Typically, a larger $N$ will result in a stronger approximation, but this increases the complexity of the optimization. This tradeoff will be discussed further below. We are now prepared to discuss the optimization process. While many optimization techniques are applicable to the current problem, the technique used here will rely on iterative and procedural partial derivative descent. This will be outlined in the following section.

### 3. Procedural Partial Derivative Descent

The optimization method herein uses the gradient descent technique as a basis, which minimizes a function by systematically adjusting its parameters in a direction dictated by the opposite value of the function's gradient [17]. That is, one may minimize a function $f(x)$ iteratively by choosing an initial guess $x_0$ and executing the update algorithm:

$$\boldsymbol{x}_{p+1} = \boldsymbol{x}_p - \boldsymbol{\gamma} \cdot \nabla f(\boldsymbol{x}_p), \tag{8}$$

where $\boldsymbol{\gamma}$ is the step size vector and is chosen by the designer to facilitate the speed of convergence. Note that each element of $\boldsymbol{\gamma}$ is greater than 0. In conventional gradient descent, $\boldsymbol{\gamma}$ is instead a scalar with an optimal value determined via a line search at every step. The optimal value is determined by comparing the gradients at $\boldsymbol{x}_p$ and $\boldsymbol{x}_{p+1}$ and stopping the search when these two gradients are orthogonal [17]. If the gradient is not particularly well-behaved or if one does not wish to conduct a search at each iteration, static values or alternative adaptive strategies may be used. In either case, this value must be chosen with care. If the value is too big, the algorithm may enter an unstable trajectory and fail to converge properly. If the value is too small, convergence may take excessively long and convergence on local minima becomes more likely. We will discuss this step size further below. Our first step is to establish an appropriate $f(x)$ for the impedance approximation problem. Consider an impedance $Z_0(\omega)$ defined on the interval $\omega_1 \leq \omega \leq \omega_2$ that we wish to approximate with an impedance $Z_F(\omega)$ described by a Form I FCN. We define a mean squared error parameter $\varepsilon$ as follows:

$$\varepsilon = \frac{1}{\omega_2 - \omega_1} \int_{\omega_1}^{\omega_2} |Z_0(\omega) - Z_F(\omega)|^2 d\omega. \tag{9}$$

This is the function we wish to minimize over the prescribed frequency interval. Traditional gradient descent experiences difficulty minimizing this function, however, due to the poor conditionality of its gradient. To demonstrate this, we express the gradient of $\varepsilon$ as follows:

$$\nabla \varepsilon = \frac{1}{\omega_2 - \omega_1} \int_{\omega_1}^{\omega_2} -2[Re(Z_0(\omega)) - Re(Z_F(\omega))]\nabla[Re(Z_F(\omega)] - 2[Im(Z_0(\omega)) - Im(Z_F(\omega))]\nabla[Im(Z_F(\omega)]\, d\omega. \tag{10}$$

Each parallel resonator has three free parameters; $R_n$, $Q_n$, and $\omega_{rn}$, whereas the series resonator is described by $R_s$, $Q_s$, and $\omega_{rs}$. We evaluate the gradients in the equation above by taking the partial derivatives of the superposed impedance expression with respect to each free variable. Recall that the impedance of the $n^{th}$ parallel resonator in the network may be expressed as

$$Z(\omega) = \frac{R_n}{1 - jQ_n\left(\frac{\omega_{rn}}{\omega} - \frac{\omega}{\omega_{rn}}\right)}. \tag{11}$$

Using eqn. (4) and eqn. (11), we obtain the following partial derivatives:

$$\frac{\partial Z_F(\omega)}{\partial R_n} = \frac{1}{1 + Q_n^2\left(\frac{\omega_{rn}}{\omega} - \frac{\omega}{\omega_{rn}}\right)^2} + j\frac{Q_n\left(\frac{\omega_{rn}}{\omega} - \frac{\omega}{\omega_{rn}}\right)}{1 + Q_n^2\left(\frac{\omega_{rn}}{\omega} - \frac{\omega}{\omega_{rn}}\right)^2}, \tag{12a}$$

$$\frac{\partial Z_F(\omega)}{\partial Q_n} = \frac{-2R_n Q_n \left(\frac{\omega_{rn}}{\omega} - \frac{\omega}{\omega_{rn}}\right)^2}{\left[1 + Q_n^2\left(\frac{\omega_{rn}}{\omega} - \frac{\omega}{\omega_{rn}}\right)^2\right]^2} + j\frac{\left[\left(\frac{\omega_{rn}}{\omega} - \frac{\omega}{\omega_{rn}}\right) - Q_n^2\left(\frac{\omega_{rn}}{\omega} - \frac{\omega}{\omega_{rn}}\right)^3\right]}{\left[1 + Q_n^2\left(\frac{\omega_{rn}}{\omega} - \frac{\omega}{\omega_{rn}}\right)^2\right]^2}, \tag{12b}$$

$$\frac{\partial Z_F(\omega)}{\partial \omega_{rn}} = \frac{-2R_n Q_n^2\left(\frac{1}{\omega} + \frac{\omega}{\omega_{rn}^2}\right)\left(\frac{\omega_{rn}}{\omega} - \frac{\omega}{\omega_{rn}}\right)}{\left[1 - Q_n^2\left(\frac{\omega_{rn}}{\omega} - \frac{\omega}{\omega_{rn}}\right)^2\right]^2 + 4Q_n^2\left(\frac{\omega_{rn}}{\omega} - \frac{\omega}{\omega_{rn}}\right)^2} + j\frac{R_n Q_n\left(\frac{1}{\omega} + \frac{\omega}{\omega_{rn}^2}\right)\left[1 - Q_n^2\left(\frac{\omega_{rn}}{\omega} - \frac{\omega}{\omega_{rn}}\right)^2\right]}{\left[1 - Q_n^2\left(\frac{\omega_{rn}}{\omega} - \frac{\omega}{\omega_{rn}}\right)^2\right]^2 + 4Q_n^2\left(\frac{\omega_{rn}}{\omega} - \frac{\omega}{\omega_{rn}}\right)^2}, \tag{12c}$$

$$\frac{\partial Z_F(\omega)}{\partial R_s} = 1 - jQ_s\left(\frac{\omega_{rs}}{\omega} - \frac{\omega}{\omega_{rs}}\right), \tag{12d}$$

$$\frac{\partial Z_F(\omega)}{\partial Q_s} = -jR_s\left(\frac{\omega_{rs}}{\omega} - \frac{\omega}{\omega_{rs}}\right), \tag{12e}$$

$$\frac{\partial Z_F(\omega)}{\partial \omega_{rs}} = -jR_s Q_s\left(\frac{1}{\omega} + \frac{\omega}{\omega_{rs}^2}\right). \tag{12f}$$

Note the above expressions are broken up into their real and imaginary parts since, in this case,

$$\nabla[Re(Z_F(\omega)] = Re(\nabla[Z_F(\omega)]), \tag{13a}$$

$$\nabla[Im(Z_F(\omega)] = Im(\nabla[Z_F(\omega)]). \tag{13b}$$

While not especially apparent by inspection, the partial derivatives with respect to the $\omega_r$ parameters in both the series and parallel cases are many orders of magnitude smaller than those with respect to the $R$ and $Q$ values. This is due to the $\left(\frac{1}{\omega} + \frac{\omega}{\omega_{rn}^2}\right)$ terms present in the $\omega_r$ partial derivative expressions, which are far smaller than the $\left(\frac{\omega_{rn}}{\omega} - \frac{\omega}{\omega_{rn}}\right)$ terms in the other derivatives for high frequencies (note that accelerator impedances are routinely computed well into the GHz range). This means that the step size vector components that correspond to the $\omega_r$ parameters must be far larger than the other components, otherwise the resonant frequencies will barely change from update to update. This makes the step size vector problem-dependent in a manner that is challenging to generalize. To overcome this problem, we will modify the gradient descent technique by updating parameters individually instead of simultaneously. Further, we will apply a procedural component to the algorithm and limit which parameters can be modified under which circumstances. Specifically, the parameters belonging to each individual resonator will be grouped together. The algorithm will only update the parameters of one resonator at a time, keeping the parameters of all other resonators static. Furthermore, the algorithm will prioritize updating the active resonator's resonant frequency first, only updating the resistance or Q parameters if changing the resonant frequency in any direction fails to reduce the value of $\varepsilon$. The direction in which the resonant frequency is updated is dependent on the corresponding partial derivative of $\varepsilon$. If the partial derivative is positive, the resonant frequency will be decreased by some static value chosen by the designer. If the derivative is negative, the resonant frequency will be increased.

When the error can no longer be decreased by modifying the resonant frequency of the active resonator, $\frac{\partial \varepsilon}{\partial R_n}$ is compared to $\frac{\partial \varepsilon}{\partial Q_n}$ ($\frac{\partial \varepsilon}{\partial R_s}$ and $\frac{\partial \varepsilon}{\partial Q_s}$ are compared if the active resonator is the series one). The parameter that corresponds to the larger partial derivative absolute value will be chosen to modify next. That is, if $\left|\frac{\partial \varepsilon}{\partial R_n}\right| > \left|\frac{\partial \varepsilon}{\partial Q_n}\right|$, then $R_n$ will be updated as follows:

$$R_n^{update} = R_n - \gamma \frac{\partial \varepsilon}{\partial R_n}. \tag{14}$$

If the converse is true, then $Q_n$ will be updated similarly:

$$Q_n^{update} = Q_n - \gamma \frac{\partial \varepsilon}{\partial Q_n}. \tag{15}$$

After the update is conducted, $\varepsilon$ is computed again using the new values to determine if the error was reduced. If not, the update is undone and then recomputed using a smaller value of $\gamma$. In this way, we are able to utilize the advantages of both large and small step sizes. At the beginning of the loop, the step size is chosen to be large to facilitate faster convergence, but it is then decreased in magnitude to better approach the true minimum of $\varepsilon$. Note that for this algorithm, the step size is shared between $R_n$ and $Q_n$, but this is not required in general. The comparison-update procedure is repeated until $\varepsilon$ cannot be reduced without decreasing the step size beyond a certain user-defined threshold or if the change in $\varepsilon$ between updates is smaller than some user-defined value. Once one of these conditions is met, the active resonator parameters are then saved and the algorithm moves to the next resonator to repeat the process. During the active resonator increment step, the algorithm also subtracts the impedance of the current resonator from target impedance (the target impedance is the impedance we are trying to develop a resonator network to approximate), minimizing the interference between the fits of the active and subsequent resonators. Once all the resonators have been updated, the code resets the target impedance to its initial value and rolls back to the first resonator to begin a new iteration of updates, once again moving from active resonator to active resonator while keeping the others static. The designer chooses how many of these update iterations occurs, but multiple iterations are necessary to ensure that $\varepsilon$ cannot be reduced further with this technique since modifying each resonator changes the impedance landscape of the entire equivalent circuit. The designer also chooses the number of resonators to include in the network, with more typically being necessary to model more complex impedances. However, if the code determines that only a subset of the resonators is needed, it typically nulls the extraneous resonators by setting their $R_n$ values to 0 (with the $Q$ value being set to 0 as well if the active resonator is the series one). The code terminates when all iterations have been completed. The process outlined above is illustrated at a high level by the flowchart in Fig. 2. Note that the resonator parameters are initialized according to the designer. As with the step size, many initialization choices are possible, but setting $Q_n$ equal to 1 and $R_n$ equal to the maximum value of the real part of the current impedance tends to work well empirically. Setting both $Q_n$ and $R_n$ equal to 10 has also yielded acceptable results. The initial $\omega_{rn}$ values can either be the frequencies where the $R_n$ maxima reside or they can be generated randomly relative to a uniform distribution depending on the problem at hand. The number of resonators to include is also up to the designer. Obviously, including more resonators will increase computation time, but this can result in a more sophisticated network frequency response if the target impedance profile has many features. Within reason, it is prudent to err on the side of more resonators instead of fewer, with the current study using 100 resonator networks as a base. It should be noted that the algorithm will not necessarily use all 100 resonators. When this happens, the algorithm, on its own, will only modify a subset of the resonators from iteration to iteration, nulling the impact of the extraneous resonators. Empirically, this is the typical result, meaning that adding more resonators gives rise to a point of diminishing returns for most problems. Where that point is exactly, however, is dependent on the problem itself and some experimentation may be needed. Nevertheless, 100 resonators as a general conservative starting point is empirically a tenable choice for consistent results. Furthermore, the time spent on extraneous resonators is comparatively minimal since a loop where no possible changes result in an $\varepsilon$ reduction terminates much faster than a loop which moves the network many steps toward the $\varepsilon$ minimization.

Before moving on, it should also be discussed whether the algorithm should be allowed to return negative values for $R_n$, $Q_n$, or $\omega_{rn}$. While having such parameters in a real resonator would be nonphysical, this lack of physicality does not propagate to the approximated wake potential if the excitation current and target impedance profile are physical and all resonator parameters are real. There is quantitative basis for this claim. A nonphysical wake potential would be one with a nonzero imaginary part (negative wake potential values are allowed).

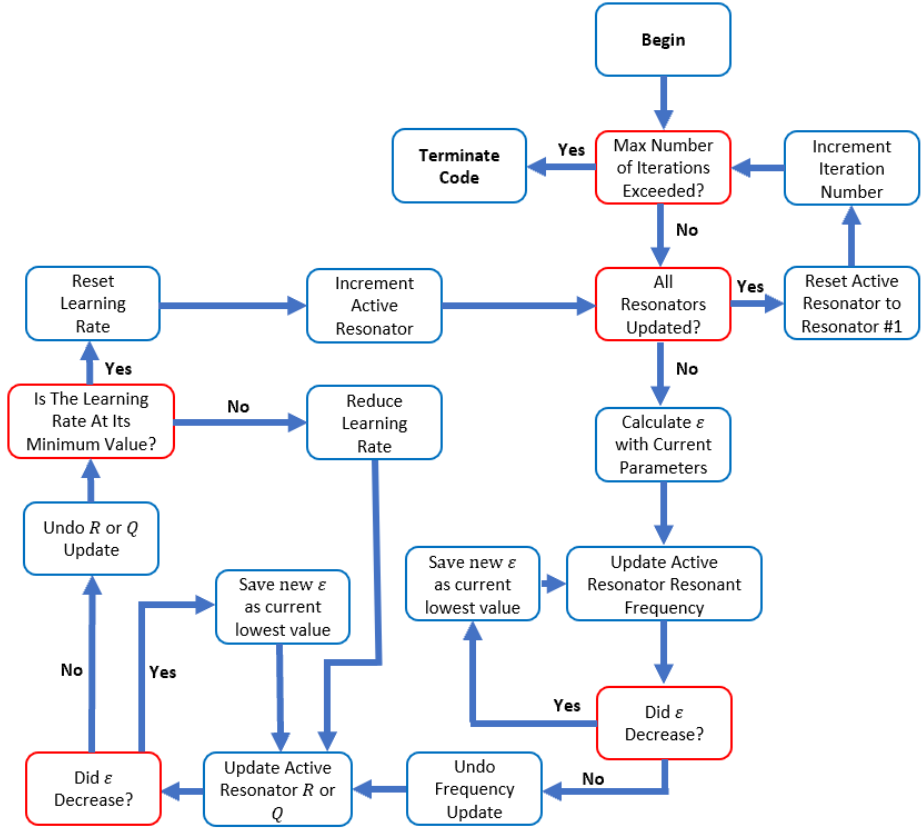

**FIG. 2.** High-level flowchart of partial derivative descent algorithm.

Recalling eqn. (7), we see that if the $R$ values are real, there is no way for them to introduce an imaginary part regardless of their sign. The analysis of the effect of the $Q$ and $\omega_r$ values, on the other hand, is less obvious, but still straightforward. The $Q$ parameters appear both explicitly and within the $r^\pm$ values along with the $\omega_r$ parameters. The explicit $Q$ contributions, if real, do not cause issues for the same reasons the $R$ contributions don't. To conclude that the $r^\pm$ values also do not cause problems, we observe that $r^+$ and $r^-$ have the same imaginary part and real parts of opposite sign. This means that $\tilde{I}(r_n^+)$ and $\tilde{I}(r_n^-)$ are complex conjugates of each other. To see why, we invoke the following Fourier Transform Property:

$$\check{F}(f(t)e^{j\omega_o t}) = F(\omega - \omega_0). \tag{16}$$

This implies:

$$\check{F}\left(I(t)e^{j(\frac{-j\omega_r}{2Q})t}\right)\bigg|_{\omega=\pm\frac{\omega_r}{2}\sqrt{4-\frac{1}{Q^2}}} = \int_{-\infty}^{\infty} I(t)e^{j(\frac{-j\omega_r}{2Q})t}e^{\pm j\frac{\omega_r}{2}\sqrt{4-\frac{1}{Q^2}}t}dt = \tilde{I}\left(\pm\frac{\omega_r}{2}\sqrt{4-\frac{1}{Q^2}} - \frac{-j\omega_r}{2Q}\right) = \tilde{I}(r^\pm). \tag{17}$$

From Euler's Formula, we can expand the Fourier Integral:

$$\int_{-\infty}^{\infty} I(t)e^{j(\frac{-j\omega_r}{2Q})t}e^{\pm j\frac{\omega_r}{2}\sqrt{4-\frac{1}{Q^2}}t}dt = \int_{-\infty}^{\infty} I(t)e^{\frac{\omega_r t}{2Q}}\cos\left(\frac{\omega_r}{2}\sqrt{4-\frac{1}{Q^2}}t\right)dt \pm j\int_{-\infty}^{\infty} I(t)e^{\frac{\omega_r t}{2Q}}\sin\left(\frac{\omega_r}{2}\sqrt{4-\frac{1}{Q^2}}t\right)dt. \tag{18}$$

Therefore, $\tilde{I}(r_n^+)$ and $\tilde{I}(r_n^-)$ are complex conjugates. By expanding terms, it is straightforward to see that $e^{j\omega_{rn}r_n^+ t}$ and $e^{j\omega_{rn}r_n^- t}$ are also complex conjugates, meaning $\tilde{I}(r_n^+)e^{j\omega_{rn}r_n^+ t}$ and $\tilde{I}(r_n^-)e^{j\omega_{rn}r_n^- t}$ are as well. Now, let

$$r^+ = -j\frac{\omega_r}{2}\left(-\frac{1}{Q} + j\sqrt{4-\frac{1}{Q^2}}\right) \equiv -jr_c, \tag{19a}$$

which means

$$r^- = -j\frac{\omega_r}{2}\left(-\frac{1}{Q} - j\sqrt{4 - \frac{1}{Q^2}}\right) \equiv -j\overline{r_c}. \tag{19b}$$

We now have

$$r^+\tilde{I}(r^+)e^{jr^+t} - r^-\tilde{I}(r^-)e^{jr^-t} = -j\left(r_c\tilde{I}(r^+)e^{jr^+t} - \overline{r_c}\tilde{I}(r^-)e^{jr^-t}\right). \tag{20}$$

The terms in the parentheses represent the subtraction of some quantity from its complex conjugate, which results in a purely imaginary value. Since this is then multiplied by $-j$, we see that the expression in eqn. (20) is purely real regardless of the sign of $Q$ or $\omega_r$, provided they are both real. Finally, we note that since $r^+$ and $r^-$ have the same imaginary part, $r^+ - r^-$ is purely real as well. Therefore, the wake potential given by eqn. (7) remains real even if the $Q$ or $\omega_r$ of any of the resonators takes a negative value. In the next section, the results of applying the above algorithm to several accelerator impedance studies will be shown.

## 4. Results

To evaluate the efficacy of the presented technique, we will now map FCN's to three real-life accelerator impedance profiles across various frequency ranges using an implementation of the previously described algorithm in MATLAB. The first profile we will evaluate is the longitudinal impedance of the Lamberston magnets currently being used in the Fermilab Main Injector Ring [18]. The impedance given in [18] is shown in Fig. 3 along with the impedance of the mapped FCN synthesized by the algorithm presented above. Additionally, the resonator parameter outputs for this are given in Appendix I, as well as the code runtime to generate the network. The initial resonant frequencies for this example were generated based on real part impedance maxima as discussed in the previous section.

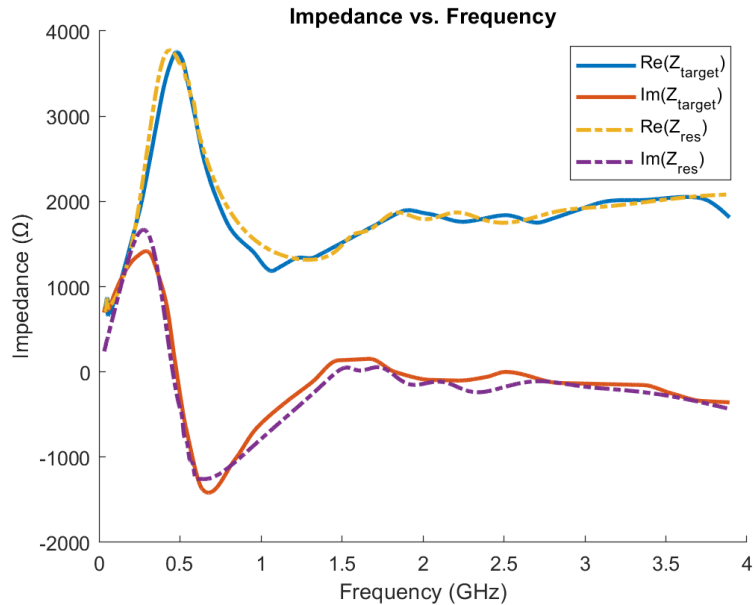

**FIG. 3.** Fermilab Main Injector Lamberston magnet and approximate resonator network impedances. The solid lines are the target impedance values and the dotted lines represent the impedance of the synthesized resonator network.

Fig. 3 shows very good agreement between the synthesized network impedance and the true values, indicating, at a minimum, that the technique presented here is appropriate for smooth broadband impedance profiles with few stark features. This network consists of 1 series resonator and 16 parallel resonators with various parameters and was generated over 5 iterations.

Next, we will consider the total broadband longitudinal impedance of the NCLS-II Storage Ring for a .3 mm bunch length, which is given in [19]. This impedance is given over a frequency span of 300 GHz and requires 30 parallel resonators to model. As with Fig. 3, the comparison between the target and synthesized network impedances for this case are given in Fig. 4. As with the Lamberston magnet example, the initial resonant frequencies used the locations of the real part impedance maxima as a starting point.

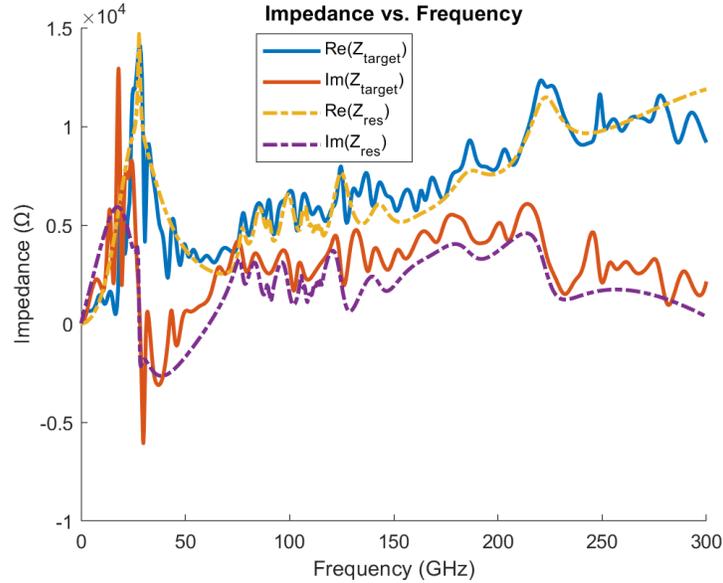

**FIG. 4.** NCLS-II Storage Ring and approximate resonator network impedances. The solid lines are the target impedance values and the dotted lines represent the impedance of the synthesized resonator network.

Here, the agreement between the synthesized and target impedance profiles is lower than with the Lamberston magnet case, which is expected since the frequency range is much wider with far more features. For instance, there is a spike in the imaginary part of the target impedance at around 20 GHz that is not captured by the resonator network. Additionally, there is a dip in the real part of the target impedance around 30 GHz that is also not captured. Nevertheless, the resonator network impedance follows the target impedance quite strongly overall, certainly well enough for first-order simulation studies.

Finally, we will consider the longitudinal impedance of a stripline kicker with two 90° electrodes, the impedance of which is given in [19] and [20]. As before, we compare this impedance to a synthesized resonator network impedance in Fig. 5. With this example, the initial resonant frequencies were generated randomly, as this yielded better results.

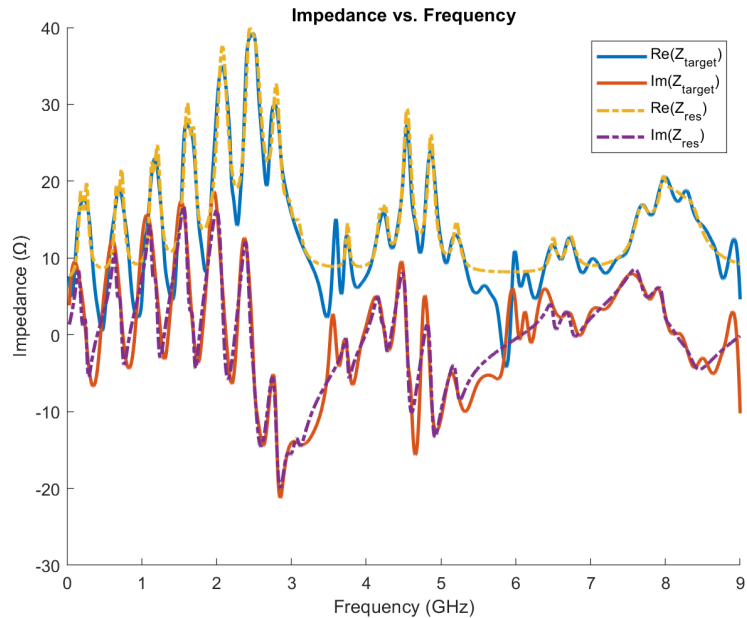

**FIG. 5.** Stripline kicker and approximate resonator network impedances. The solid lines are the target impedance values and the dotted lines represent the impedance of the synthesized resonator network.

Here, once more, we see good agreement between the synthesized network impedance and the target, though some of the oscillatory features in the real part of the impedance are not fully captured before 3 GHz. Nevertheless, the resonator network result outperforms the theoretical approximation of the kicker impedance at frequencies below 3 GHz, which is shown alongside the numerically calculated impedance in Fig. 6 of [19]. This indicates that the synthesized network impedance is an appropriate approximation since it matches the broadband target impedance well and constitutes an overall improvement over a theoretical approximation within the frequency range where the theoretical treatment is valid. Note that the algorithm for this network, as opposed to the previous examples, used all 100 resonators to properly capture the asymmetric oscillations and other features. It should also be noted that several values of $R_n$ and $Q_n$ for this network are negative. This is acceptable given the discussion in the previous section.

## 5. Conclusion

In this work, we have developed a method to represent general real-life accelerator component impedance profiles in terms of synthesized RLC resonator networks using partial derivative descent optimization. We have applied this technique to three broadband frequency responses relevant to the accelerator field and have shown solid agreement between these profiles and the impedances of the synthesized networks. This technique is highly useful since the wake potentials of such resonator networks may be computed analytically in a straightforward manner, meaning that, once the network is generated, the components' wake potentials may be known at all time points without the need for repetitive convolution, saving significant time and computing resources during beamline simulations. As such, the work presented here can be used to great effect during accelerator design efforts.

## Appendix I

Below are tables listing the resonator parameters of the networks synthesized in the three examples of Section 4. Note that, if a table contains fewer than 100 resonators, then the algorithm returned null values for the resonators not included, indicating their lack of necessity. Note also that Resonator #1 for all the networks is the series RLC. Each table caption also contains the algorithm runtime for each example on a generic laptop with 32 GB of RAM as well as the number of iterations completed.

| Resonator Number | R (Ω) | Q | $f_r$ (GHz) |
| --- | --- | --- | --- |
| 1 | 695.6189271 | 0 | 4 |
| 2 | 3061.165433 | 1 | 0.436 |
| 3 | 1311.930262 | 1 | 4.082 |
| 4 | 390.5001476 | 4.712186274 | 2.21 |
| 5 | 181.4306801 | 3.965788261 | 2.842 |
| 6 | 536.217187 | 4.543105508 | 1.826 |
| 7 | 160.2546347 | 24.77271266 | 0.586 |
| 8 | 163.11532 | 24.29191781 | 0.552 |
| 9 | 126.5849183 | 1316.866073 | 0.05 |
| 10 | 125.7083409 | 1320.051448 | 0.049 |
| 11 | 122.2372349 | 1274.958927 | 0.051 |
| 12 | 120.6887107 | 1237.364837 | 0.048 |
| 13 | 112.1013301 | 1234.916788 | 0.047 |
| 14 | 112.4928412 | 1198.894385 | 0.052 |
| 15 | 168.5624404 | 24.50549518 | 0.519 |
| 16 | 215.8332079 | 7.451141111 | 1.568 |
| 17 | 94.50700975 | 1063.261728 | 0.053 |

**Table A.1.** Resonator parameters for the network synthesized to approximate the Fermilab Main Injector Lamberston magnet longitudinal impedance. 5 iterations. Runtime = 3 min.

| Resonator Number | R (Ω) | Q | $f_r$ (GHz) |
|---|---|---|---|
| 1 | 0 | 0 | 300 |
| 2 | 5198.1451 | 21.2441703 | 27.832 |
| 3 | 12337.9601 | 1 | 339.61 |
| 4 | 9452.64631 | 1 | 26.787 |
| 5 | 4666.20179 | 11 | 124.65 |
| 6 | 4025.13104 | 11 | 99.261 |
| 7 | 3382.48986 | 11 | 85.623 |
| 8 | 2094.21066 | 10.9884028 | 142.84 |
| 9 | 1919.37856 | 46.7412581 | 107.8 |
| 10 | 2139.06261 | 18.0826521 | 77.345 |
| 11 | 2335.23576 | 13.5567097 | 222.48 |
| 12 | 1572.74661 | 7.43761208 | 221.41 |
| 13 | 1207.84189 | 9.45003197 | 186.03 |
| 14 | 834.751323 | 148.202949 | 89.991 |
| 15 | 983.598468 | 50.4323558 | 110.17 |
| 16 | 598.720206 | 120.517882 | 91.088 |
| 17 | 912.363757 | 66.6852037 | 112.28 |
| 18 | 862.039661 | 5.3971679 | 185.16 |
| 19 | 1202.52388 | 80.6345988 | 106.09 |
| 20 | 755.805952 | 92.5530311 | 115.05 |
| 21 | 480.60856 | 234.720583 | 90.624 |
| 22 | 888.477338 | 5.01303663 | 163.09 |
| 23 | 522.098108 | 166.148267 | 89.357 |
| 24 | 661.998875 | 123.95807 | 113.96 |
| 25 | 298.293384 | 541 | 90.352 |
| 26 | 419.748066 | 187.417967 | 113.25 |
| 27 | 90.9196336 | 691 | 90.472 |
| 28 | 80.1254642 | 721 | 90.234 |
| 29 | 0.01711689 | 691 | 90.235 |
| 30 | 0.00233523 | 691 | 90.235 |
| 31 | 10.0003186 | 691 | 90.474 |

**Table A.2.** Resonator parameters for the network synthesized to approximate the NSLS-II total longitudinal impedance. 5 iterations. Runtime = 70 min.

| Resonator Number | R (Ω) | Q | $f_r$ (GHz) |
|---|---|---|---|
| 1 | 12.19772911 | 1.11323568 | 0.225021523 |
| 2 | 19.22202327 | 2.934148588 | 0.698074272 |
| 3 | 13.92773618 | 3.656775963 | 1.176127579 |
| 4 | 23.63309888 | 3.849721156 | 1.596174417 |
| 5 | 29.06199253 | 7.549918617 | 2.074227724 |
| 6 | 37.34173614 | 10.54813445 | 2.487273782 |
| 7 | 12.81716377 | 3.849702736 | 2.774305788 |
| 8 | 15.08488578 | 30.07010268 | 3.596397457 |
| 9 | 12.98892641 | 19.57860456 | 3.742413739 |
| 10 | 15.92137084 | 17.82554248 | 4.230468161 |
| 11 | 27.33505342 | 34.20398723 | 4.546503401 |
| 12 | 23.99433913 | 31.03817102 | 4.863538753 |
| 13 | -18.40020959 | 2.351887848 | 5.191575332 |
| 14 | 0.585368187 | -2.916066794 | 5.579618602 |
| 15 | 10.935892 | 24.79179408 | 5.987664102 |
| 16 | 8.157196511 | 29.47475557 | 6.139681053 |
| 17 | 11.76482029 | 2.255767743 | 6.492720419 |
| 18 | -67.33059102 | 0.1453626 | 6.718745623 |
| 19 | -0.22842353 | -7.221489971 | 7.030780417 |
| 20 | -12.62646613 | 5.999752737 | 7.693854355 |
| 21 | 17.29887078 | 17.08971109 | 7.996888145 |
| 22 | 14.87008351 | 13.68732476 | 8.277919483 |
| 23 | 10 | 10 | 8.149905208 |
| 24 | 0.936950854 | 10 | 7.891876436 |
| 25 | 10 | 10 | 8.998999888 |
| 26 | 0.379444693 | 10 | 7.78286428 |
| 27 | 10 | 10 | 0.362036802 |
| 28 | -6.548679649 | 10 | 4.916544664 |
| 29 | 10 | 10 | 8.978997658 |
| 30 | 3.910847888 | 10 | 4.614510985 |
| 31 | 1.64220667 | 10 | 7.865873536 |
| 32 | 10 | 10 | 0.661070146 |
| 33 | 10 | 0.361000484 | 8.888987621 |
| 34 | -4.16752294 | 10 | 8.306922717 |
| 35 | 8.184893978 | 10 | 5.092564291 |
| 36 | 7.972849473 | 10 | 3.901431471 |
| 37 | 6.070945386 | 10 | 3.061337794 |
| 38 | 10 | 10 | 6.495720754 |
| 39 | 10 | 10 | 0.154013605 |
| 40 | 4.892000011 | 10 | 3.386374038 |
| 41 | 10 | 10 | 8.306922717 |
| 42 | -0.572404028 | 10 | 4.932546448 |
| 43 | 0.401245548 | 10 | 4.281473849 |
| 44 | -3.282140008 | 10 | 4.484496487 |
| 45 | 10 | 10 | 2.80230891 |
| 46 | 10 | 0.330756392 | 8.559950931 |
| 47 | 10 | 0.359070334 | 8.838982045 |
| 48 | 2.38235016 | 10 | 4.63751355 |
| 49 | -15.46477073 | 10 | 8.93399264 |
| 50 | 6.064634346 | 10 | 4.120455894 |
| 51 | -4.912296784 | 10 | 3.852426006 |
| 52 | -16.87721086 | 5.618030212 | 1.944213226 |
| 53 | 10 | 10 | 1.765193264 |
| 54 | 10 | 0.141239052 | 7.500832831 |
| 55 | 10 | 3.674841674 | 6.548726664 |
| 56 | -1.36938276 | 10 | 4.78252972 |
| 57 | 10 | 0.130577679 | 7.467829151 |
| 58 | 8.919671577 | 10 | 4.622511877 |
| 59 | 4.523544116 | 10 | 4.981551913 |
| 60 | 10 | 10 | 1.944213226 |
| 61 | 10 | 10 | 5.303587822 |
| 62 | 5.853504257 | 10 | 1.312142746 |
| 63 | 2.604038227 | 10 | 0.500052191 |
| 64 | 10 | 10 | 6.159683283 |
| 65 | 9.107723561 | 10 | 5.489608565 |
| 66 | 10 | 10 | 2.001219583 |
| 67 | -7.642896794 | 10 | 3.675406267 |
| 68 | 10 | 10 | 5.681629977 |
| 69 | 10 | 10 | 5.011555258 |
| 70 | 10 | 10 | 1.176127579 |
| 71 | 10 | 10 | 1.549169176 |
| 72 | 10 | 10 | 0.040000892 |
| 73 | 4.885590023 | 10 | 3.7824182 |
| 74 | 2.838449172 | 10 | 4.412488458 |
| 75 | -8.904057209 | 10 | 1.465159808 |
| 76 | -44.90503725 | 1.359428645 | 6.01266689 |
| 77 | 10 | 10 | 0.192017843 |
| 78 | 10 | 10 | 1.105119661 |
| 79 | 10 | 0.079877003 | 8.570952158 |
| 80 | 10 | 0.361213449 | 8.783975912 |
| 81 | 10 | 10 | 0.309030891 |
| 82 | 4.469725594 | 10 | 4.461493922 |
| 83 | 2.857975316 | 10 | 7.768862719 |
| 84 | 10 | 10 | 2.210242891 |
| 85 | 10 | 0.192921078 | 7.513834281 |
| 86 | 10 | 4.2694811 | 7.32881365 |
| 87 | 10 | 10 | 5.672628973 |
| 88 | 10 | 10 | 0.05200223 |
| 89 | -12.25198492 | 10 | 3.437379726 |
| 90 | -28.69527147 | 9.918168535 | 8.142904427 |
| 91 | -21.3520451 | 10 | 6.133680384 |
| 92 | 8.291922759 | 10 | 3.429378834 |
| 93 | -8.120286858 | 10 | 5.699631984 |
| 94 | -3.707707069 | 10 | 2.213243225 |
| 95 | 8.967940828 | 10 | 5.156571429 |
| 96 | 10 | 0.365411268 | 8.836981822 |
| 97 | 8.518395337 | 10 | 7.651849671 |
| 98 | -2.322655242 | 10 | 2.573283372 |
| 99 | 8.831817142 | 10 | 6.152682503 |
| 100 | 4.127278056 | 10 | 3.243358091 |

**Table A.3.** Resonator parameters for the network synthesized to approximate the stripline kicker total longitudinal impedance. 5 iterations. Runtime = 53 min.